\documentclass[10pt, conference]{IEEEtran}
\IEEEoverridecommandlockouts
 \usepackage{cite}
\usepackage{amsmath,amssymb,amsfonts}
\usepackage{algorithmic}
\usepackage{graphicx}
\usepackage{textcomp}
\usepackage{xcolor}
\usepackage{soul}
\usepackage{ulem}
\usepackage{listings}
\usepackage{xcolor}
\usepackage[utf8]{inputenc}
\usepackage{pifont}
\usepackage{xcolor}
\newcommand{\cmark}{\textcolor{green!80!black}{\ding{51}}}

\usepackage{makecell}
\usepackage{multirow}
\newcommand{\tabincell}[2]{\begin{tabular}{@{}#1@{}}#2\end{tabular}}

\definecolor{codegreen}{rgb}{0,0.6,0}
\definecolor{codegray}{rgb}{0.5,0.5,0.5}
\definecolor{codepurple}{rgb}{0.58,0,0.82}
\definecolor{backcolour}{rgb}{0.95,0.95,0.92}

\lstdefinestyle{mystyle}{
    backgroundcolor=\color{backcolour},   
    commentstyle=\color{codegreen},
    keywordstyle=\color{magenta},
    numberstyle=\tiny\color{codegray},
    stringstyle=\color{codepurple},
    basicstyle=\ttfamily\footnotesize,
    breakatwhitespace=false,         
    breaklines=true,                 
    captionpos=b,                    
    keepspaces=true,                 
    numbers=left,                    
    numbersep=5pt,                  
    showspaces=false,                
    showstringspaces=false,
    showtabs=false,                  
    tabsize=2,
    float=tp,
  floatplacement=tbp,
  abovecaptionskip=-2pt
}

\def\BibTeX{{\rm B\kern-.05em{\sc i\kern-.025em b}\kern-.08em
    T\kern-.1667em\lower.7ex\hbox{E}\kern-.125emX}}
\begin{document}

\title{On Runtime Software Security of TrustZone-M based IoT Devices
}

    

\author{\IEEEauthorblockN{Lan Luo}
\IEEEauthorblockA{\textit{Department of Computer Science} \\
\textit{University of Central Florida}\\
Orlando, USA \\
lukachan@knights.ucf.edu}
\and
\IEEEauthorblockN{Yue Zhang}
\IEEEauthorblockA{\textit{College of Information Science and Technology} \\
\textit{Jinan University}\\
Guangzhou, China \\
zyueinfosec@gmail.com}
\and
\IEEEauthorblockN{Cliff C. Zou}
\IEEEauthorblockA{\textit{Department of Computer Science} \\
\textit{University of Central Florida}\\
Orlando, USA \\
czou@cs.ucf.edu}
\and
\IEEEauthorblockN{Xinhui Shao}
\IEEEauthorblockA{\textit{School of Cyber Science \& Engineering} \\
\textit{Southeast University}\\
Nanjing, China \\
xinhuishao@seu.edu.cn}
\and
\IEEEauthorblockN{Zhen Ling}
\IEEEauthorblockA{\textit{School of Computer Science \& Engineering} \\
\textit{Southeast University}\\
Nanjing, China \\
zhenling@seu.edu.cn}
\and
\IEEEauthorblockN{Xinwen Fu}
\IEEEauthorblockA{\textit{Department of Computer Science} \\
\textit{University of Massachusetts Lowell}\\
Lowell, USA \\
xinwenfu@cs.uml.edu}
}

\maketitle

\begin{abstract}
Internet of Things (IoT) devices have been increasingly integrated into our daily life. However, such smart devices suffer a broad attack surface. Particularly, attacks targeting the device software at runtime are challenging to defend against if IoT devices use resource-constrained microcontrollers (MCUs). TrustZone-M, a TrustZone extension for MCUs, is an emerging security technique fortifying MCU based IoT devices.
This paper presents the first security analysis of potential software security issues in TrustZone-M enabled MCUs. We explore the stack-based buffer overflow (BOF) attack for code injection, return-oriented programming (ROP) attack, heap-based BOF attack, format string attack, and attacks against Non-secure Callable (NSC) functions in the context of TrustZone-M. 
We validate these attacks using the TrustZone-M enabled SAM L11 MCU. Strategies to mitigate these software attacks are also discussed.
\end{abstract}


\section{Introduction}
\label{sec:intro}

The Internet of Things (IoT) industry is booming, but has attracted cybercriminals \cite{sonicwall,owasp}. IoT has a broad range of application domains such as home appliances, medical instruments, industry automation, and smart buildings. It is reported that more than 20 billion IoT devices have been distributed worldwide and this number will reach 41 billions by 2027 \cite{IoTnumber}.
In this paper, we focus on IoT devices using low-cost and resource-constrained microcontrollers (MCUs), which can communicate with the outside world through venues such as WiFi, Bluetooth, NB-IoT and LoRa. The attack surface of such IoT devices includes data, networking, hardware and  software \cite{mohanty2018control, haddadpajouh2019survey, english2019exploiting, nyman2017cfi}. We are particularly interested in runtime software security of MCUs. Even if software/firmware integrity can be verified at boot time via mechanisms like secure boot, protecting software of embedded devices at runtime is challenging due to the heterogeneity and constrained computational resources of MCUs. 


TrustZone-M, the TrustZone extension for ARMv8-M architecture, is an emerging solution to the runtime software security of IoT devices \cite{tzm,liu2019security,jung2020secure}.
Specifically, it provides resource-constrained MCUs a lightweight hardware-based solution to a trusted execution environment (TEE) for security related software, i.e., Secure World (SW), isolated from the rich execution environment (REE) for the rest of the applications, i.e., Non-secure World (NSW). The NSW code cannot access the SW resources directly.
TrustZone-M provides a Non-secure Callable (NSC) memory region in the SW so that functions can be defined in the NSC region as the gateway from the NSW to the SW. 
To the best of our knowledge, TrustZone-M has not been adopted in commercial IoT products.

In this paper, we present the first security analysis of potential software security issues in TrustZone-M enabled IoT devices. 
We find that software vulnerabilities may exist in all the regions including the NSW, NSC and SWX (which is defined as the SW region excluding the NSC) of TrustZone-M. TrustZone-M is subject to the code injection attack, code reuse attack, heap-based buffer overflow attack, format string attack, and NSC specific attacks. The first four attacks can occur in the NSW, NSC and SWX. By exploiting NSC vulnerabilities, attackers can breach the security of the SW from the NSW. \looseness=-1

A number of works have been done concerning the security issues in TrustZone. 
Cerdeira et al. \cite{cerdeira2020sok} present systematization of knowledge (SoK) on the Cortex-A TrustZone security while our work focuses on the Cortex-M TrustZone. Iannillo et al. \cite{iannillo2019proposal} propose a framework for the security analysis of TrustZone-M. However, their work does not identify concrete vulnerabilities/attacks against TrustZone-M. Jung et al. \cite{jung2020secure} design a secure platform based on the Platform Security Architecture (PSA) with a brief discussion of possible attacks. Our work demonstrates five types of realistic attacks, breaching the security of TrustZone-M.


The paper makes the following major contributions:
\begin{enumerate}
\item We present the first comprehensive security analysis of the runtime software security in TrustZone-M enabled IoT devices. 
From which, potential software attacks against TrustZone-M are presented.
The SAM L11 MCU from Microchip is often used in this paper as the instance to demonstrate the principle while the methodologies used can be extended to other similar products. We validate these attacks on SAM L11 and find that even the official code examples of SAM L11 contain security vulnerabilities.

\item To defeat these software attacks, we discuss the use of control flow integrity (CFI) and point out its limitations. We present guidelines, particularly fortifying the NSC functions, for the overall system security of TrustZone-M enabled IoT devices. 
\end{enumerate}

The rest of this paper is organized as follows. We introduce the background knowledge on ARM TrustZone-M and runtime security in IoT devices in Section \ref{sec:background}. We next present the five types of practical attacks against runtime software of TrustZone-M in Section \ref{sec:pitfall}. The evaluation of the attacks is presented in Section \ref{sec:evaluation}. 
We discuss defense mechanisms in Section \ref{sec:defense} and the paper is concluded in Section \ref{sec:conclusion}. \looseness=-1

\section{Background}
\label{sec:background}
In this section, we introduce the TrustZone-M technology and runtime software security issues in IoT applications.  

\subsection{TrustZone-M}
TrustZone for Cortex-A processors (TrustZone-A) is a security technology that isolates security-critical resources (e.g., memory, peripherals) from the rich OS and applications. An ARM system on a chip (SoC) with the TrustZone extension is split into two execution environments referred as the \textbf{Secure World} (SW) and the \textbf{Non-secure World} (NSW). Software in the SW has a higher privilege and can access resources in both the SW and the NSW, while the Non-secure software is restricted to the Non-secure resources. The NSW may communicate with the SW using the monitor mode of TrustZone-A.

Recently, the TrustZone technology has been extended to the ARMv8-M architecture as TrustZone-M for some Cortex-M series processors, which are specifically optimized for resource-constrained MCUs. TrustZone-M has the SW and NSW, but differs from TrustZone-A in terms of implementation. One prominent difference is that TrustZone-M introduces a special Secure memory region named Non-secure Callable (NSC) region to provide Secure services to Non-secure software. Transition between the two worlds through the NSC is achieved by NSC function calls and returns.

\subsection{Runtime Software Security in IoT Devices}
IoT devices connect to remote servers or controllers and receive messages from them via communication venues such as WiFi, Bluetooth, and low-power wide-area network (LPWAN). We find that MCU based IoT devices are often programmed with languages such as C and C++ because they are compact, highly efficient and have the ability of direct memory control \cite{song2019sok}. Such languages provide programmers a flexible platform to interact with the low-level hardware. On the flip side, they are notoriously error-prone and daunted with security issues. Attackers may perform runtime software attacks against vulnerable IoT devices with such features. 

Runtime software attacks may hijack the program control flow by altering the control data (e.g., return address, function pointer) or change program memory by manipulating non-control data \cite{mohanty2018control}. Often in such an attack, an adversary corrupts the vulnerable memory by inputting a carefully crafted malicious payload, which eventually results in abnormal program behaviors. 
\section{Attacks against Runtime Software in TrustZone-M}
\label{sec:pitfall}

In this section, we first present the threat model on how a TrustZone-M enabled IoT device may be attacked. We then present five runtime software attacks against TrustZone-M enabled IoT devices. We often use the SAM L11 MCU as the example while the principle is the same for all TrustZone-M enabled devices. 

\begin{figure}
\centering
\includegraphics[width=0.36\textwidth]{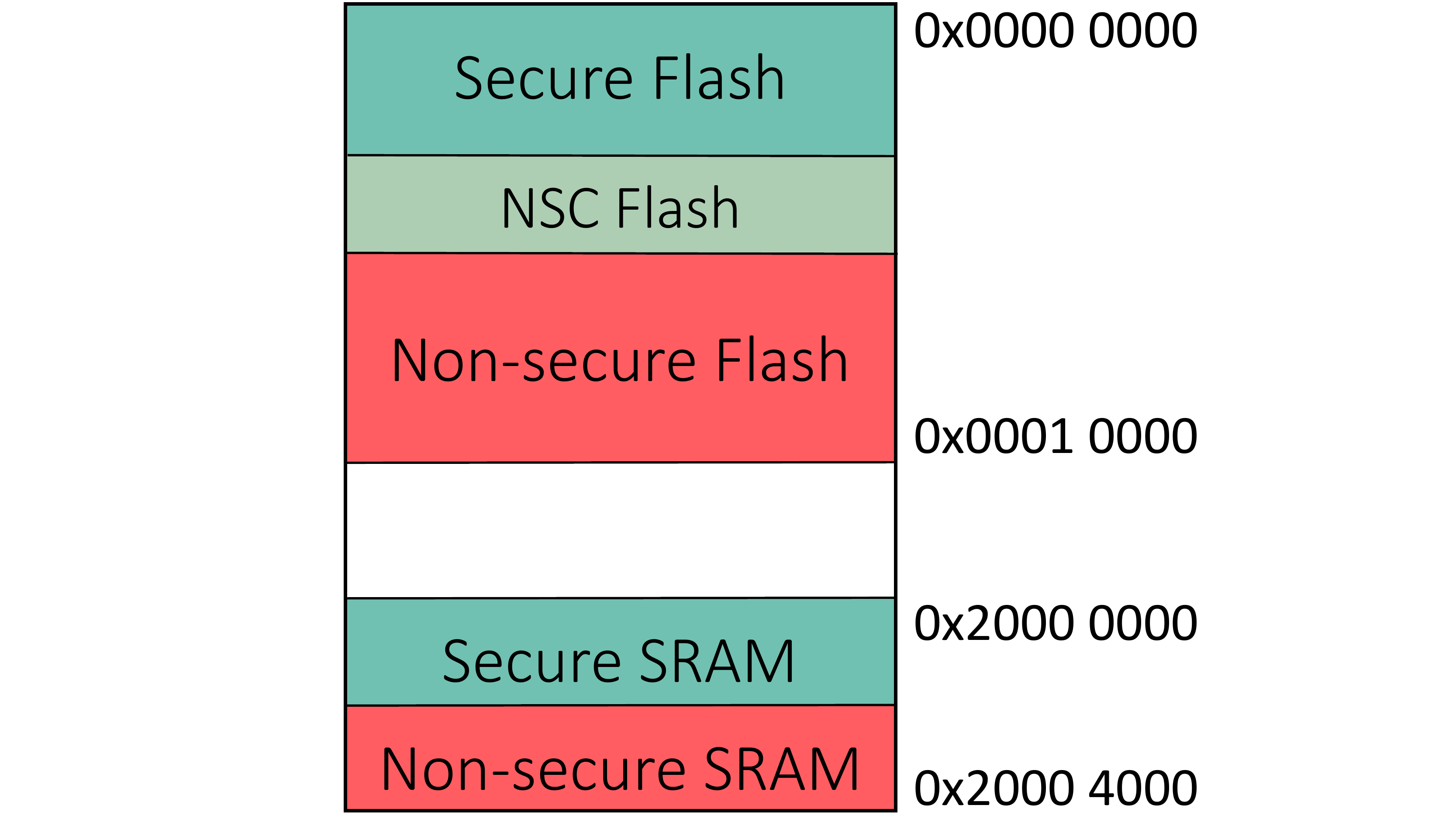}
\vspace{-2mm}
\caption{Memory layout of SAM L11. The memory is divided into the SW and NSW at the hardware level, where code in the SW (Secure Flash, NSC Flash, and Secure SRAM) can access the whole chip, while code in the NSW (Non-secure Flash and Non-secure SRAM) can only directly access resources inside the NSW.}
\label{graph::mem}
\vspace{-2mm}
\end{figure}

\subsection{Threat Model}
We consider a victim IoT device using the TrustZone-M enabled MCU. 
It is assumed that security related coding mistakes exist in the software of the victim device, which is able to receive inputs from the Internet or peripherals. Though the SW of TrustZone-M provides a TEE that the NSW software cannot directly access, the TEE can only work normally under the assumption that Secure software is well crafted with no security related coding mistakes. 
However, coding mistakes may exist in TrustZone-M's NSW, the NSC region, and the SWX region. Memory layout of a TrustZone-M based MCU, SAM L11, is shown in Figure \ref{graph::mem}.
An adversary can exploit the coding mistakes and send a malicious input (i.e. payload) to deploy software attacks. 
Even if the SW does not accept inputs from the Internet or peripherals and only the NSW communicates with the outside world, an attacker may compromise the NSW and feed malicious inputs into vulnerable NSC functions, which can access Secure resources. Therefore, if a NSC function is vulnerable, the entire SW may be compromised.

\begin{table}[t]
\centering
\caption{Software attacks in TrustZone-M} \label{table::attack}
\begin{tabular}{|l|c|c|c|} 
\hline
\textbf{Software Attacks}& \textbf{NSW}& \textbf{NSC}& \textbf{SWX}\\
\hline
Code injection& \cmark& \cmark& \cmark\\
\hline
ROP& \cmark& \cmark& \cmark\\
\hline
Heap-based BOF& \cmark& \cmark&\cmark\\
\hline
Format string attack& \cmark& \cmark& \cmark\\
\hline
NSC-specific exploit& N/A& \cmark& N/A\\
\hline
\end{tabular}
\vspace{-4mm}
\end{table}

\subsection{Runtime Software Attacks}
\label{sec:attack}

Table \ref{table::attack} lists software attacks we have identified against the NSW, NSC and SWX of TrustZone-M. It can be observed that traditional software attacks found in other platforms such as computers and smart phones can be conducted in all regions of TrustZone-M, including code injection, return-oriented programming (ROP), heap-based buffer overflow (BOF), and format string attacks. We also discover potential exploits specifically targeting the NSC.
Here, all attacks against the NSC refer to those deployed against the SW from the NSW. We present the details of these attacks in the context of TrustZone-M below.

\subsubsection{\textbf{Stack-based Buffer Overflow Attack for Code Injection}}
The stack-based buffer overflow (BOF) is a canonical memory corruption attack that occurs on the stack when a larger input is written to a local buffer without checking the buffer's boundary. Listing \ref{list::bof} presents an example, in which \textit{buf[256]} will overflow if the input is longer than 256 bytes. As a result, the extra data will overwrite the adjoining stack contents including the return address, at which the control flow will continue after the subroutine returns. Adversaries may perform the stack-based BOF attack for malicious code injection. The control flow can be redirected to the malicious code sent along with the payload by overflowing the local buffer and overwriting the original return address with the entry address of the malicious code.

\lstset{style=mystyle}
\begin{lstlisting}[language=c, caption=Example of a function with BOF vulnerability, label = list::bof] 
void BOF_func(char *input){
    char buf[256];
    strcpy(buf, input);}
\end{lstlisting}

\begin{figure}
\centering
\includegraphics[width=0.41\textwidth]{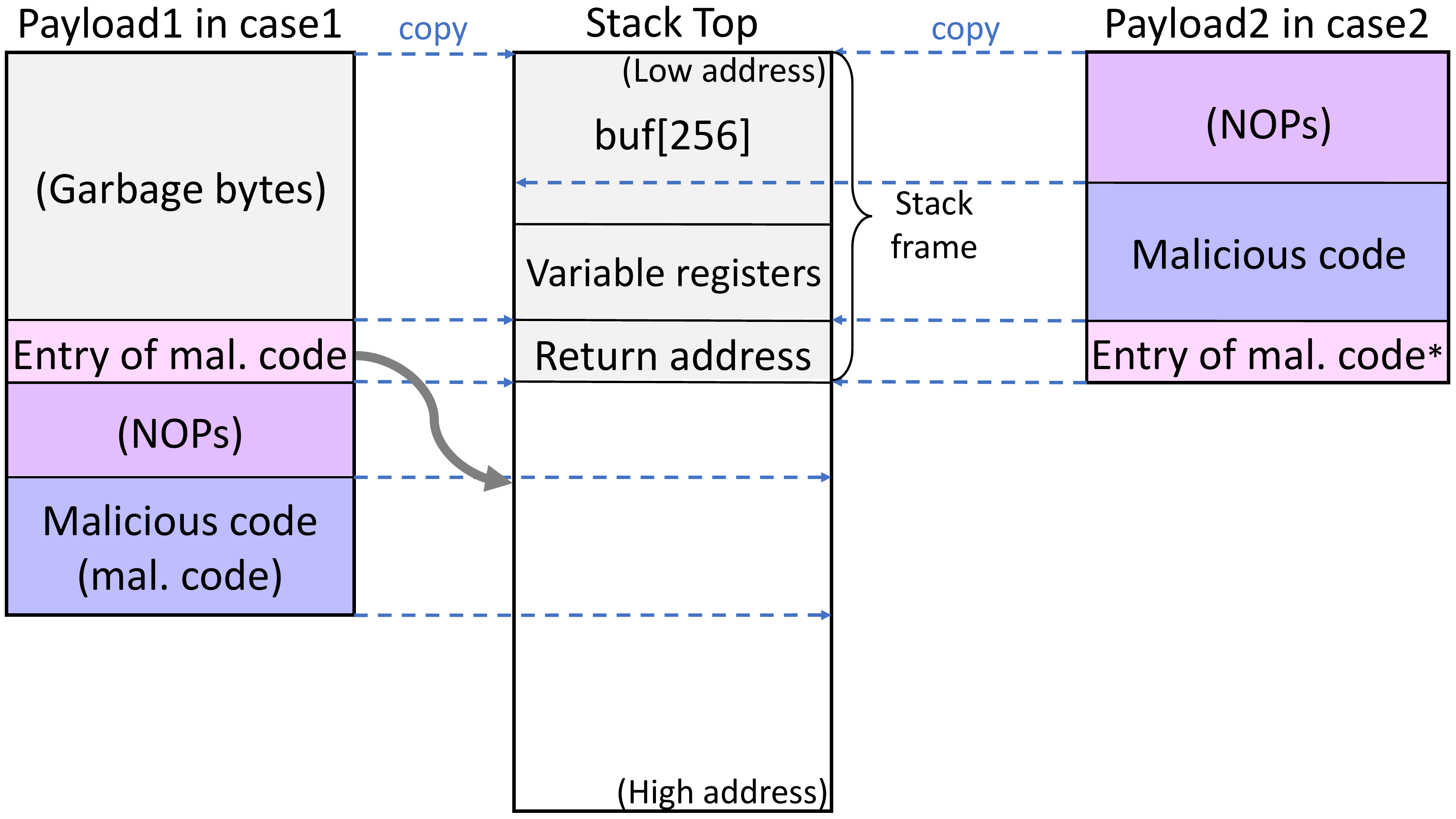}
\vspace{-2mm}
\caption{Stack-based buffer overflow attack for code injection}
\label{graph::stack}
\vspace{-4mm}
\end{figure}

To specifically implement a stack-based BOF attack against the ARMv8-M architecture, we first investigate its stack structure. A stack frame for a function in ARMv8-M consists of local variables, variable registers (R4--R7), and return address, as illustrated in Figure \ref{graph::stack}. By exploiting functions with BOF vulnerabilities, an adversary is able to copy a crafted payload to the buffer, overwrite the return address, and inject malicious code onto the stack. While constructing the malicious payload, the entry of the malicious code on the stack is needed but unknown to the adversary. A common solution is to utilize the ``JMP SP'' instruction found in the device's software\cite{rop}.
Even if there is no such instruction in the software, an adversary may enumerate possible entry addresses of malicious code to find the correct one. A wrong address in the payload lead to a program crush and restart (if automatic restart is enabled), and the malicious code would not be executed until the correct entry address is hit. This entry scanning process can be more efficient by inserting a sequence of \textit{NOP} (no-operation) instructions, called a NOP sled, before the injected code in the payload since any hit of a NOP instruction will lead to the execution of malicious code eventually.

A challenge of implementing BOF, also exists in ARMv8-M, comes from the null bytes (0x00) in the payload, which also functions as the C string terminator.
If the exploitable function treats the payload as a string (e.g., \textit{strcpy()}, \textit{strcat()}) and some null bytes exist in the crafted payload, the function will cease to copy the payload after hitting a null byte and the attack will fail. 
We discuss two cases of null bytes as follows.

First, null bytes can exist in the malicious code and the NOP sled since null bytes are naturally contained in many ARM instructions. To eliminate these null bytes, one can replace the problematic instructions by alternative instructions with the same functionalities but without null bytes. For an instance, a NOP instruction (0xBF00) can be replaced by the instruction \textit{MOV R2, R2} (0x121C).

The second case is the null bytes in the entry address of the malicious code. In SAM L11, this has to be in an area of the SRAM with a fixed address range from 0x20000000 to 0x20004000, with the upper halfword always being 0x2000, containing a null byte. A potential solution is to construct the payload like Payload 2 in Figure \ref{graph::stack}, in which the entry address is placed at the bottom. Further because the little-endian ordering in ARMv8-M, the 0x2000 is consistently located at the last two bytes of Payload 2 and shall be the only two bytes missing when copied to the stack. But, the original return address will already contain 0x2000 in its upper halfword if the caller function is executed from the SRAM, in which case the BOF will still be applicable.

\subsubsection{\textbf{Return-oriented Programming}}

BOF based code injection can be mitigated by security mechanisms like non-executable memory region \cite{nxbit}, which prevents code execution from certain memory region. However, an attacker can bypass such defense by leveraging code reuse attack (CRA). A representative CRA is return-oriented programming (ROP). Utilizing BOF to overwrite the return address, ROP redirects the control flow to a code sequence (called a gadget) found in the existing software code. 
It is also possible to chain several gadgets for more complex program control. 
Each gadget in the chain is a code segment responsible for certain operations (e.g., arithmetic operations and load/store data) and must end with the epilogue of a subroutine to help chain the gadgets. In ARMv8-M, the instruction sequence \textit{\{POP LR, BX LR\}}, which is the epilogue of leaf subroutines, pops a word to link register (\textit{LR}), and then branch to the address specified by \textit{LR}. Instruction \textit{POP PC}, which is the epilogue of non-leaf subroutines, directly pops a word to program counter (\textit{PC}).

\begin{figure}
\centering
\includegraphics[width=0.42\textwidth]{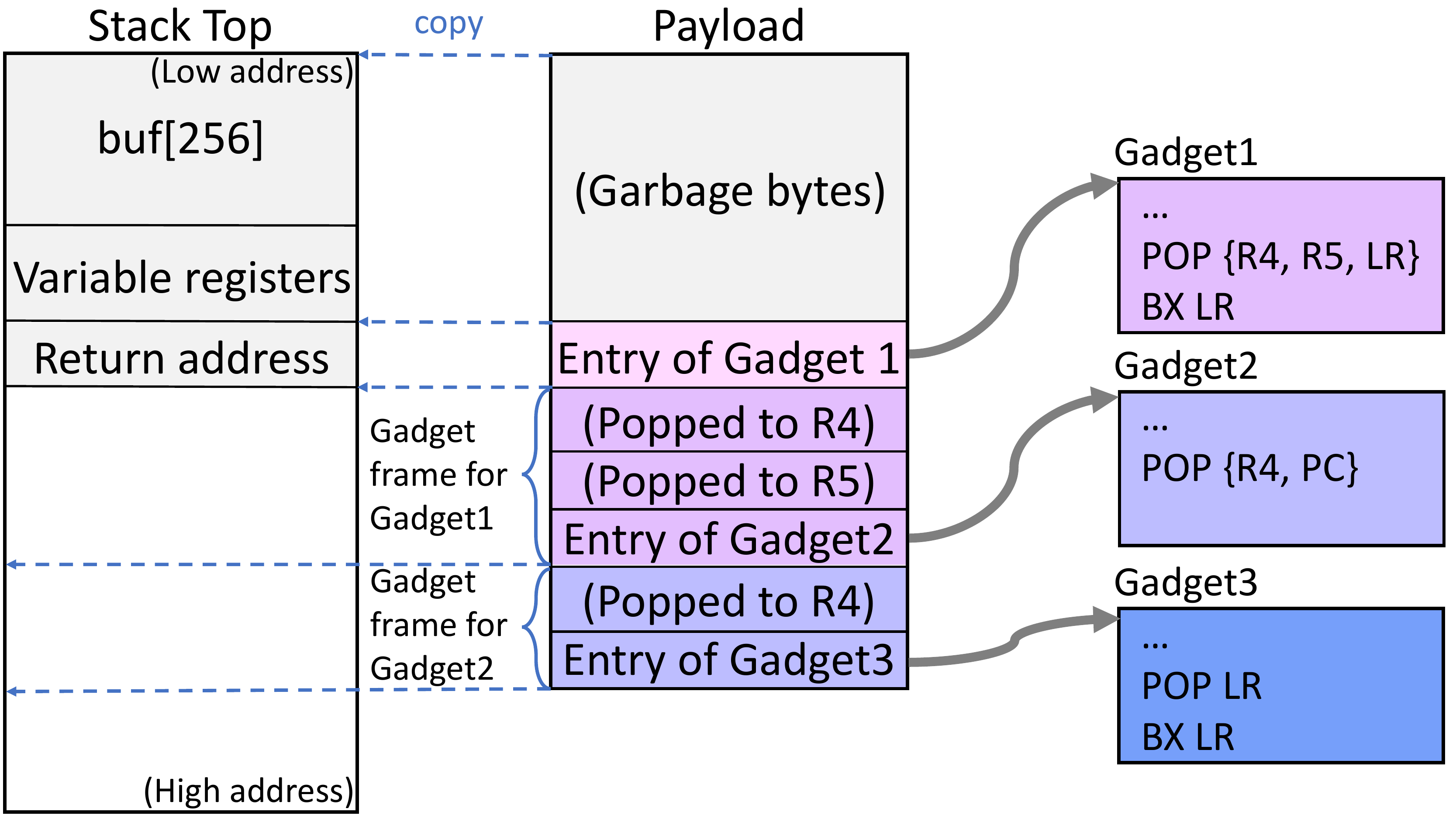}
\vspace{-2mm}
\caption{Return-oriented programming}
\label{graph::rop}
\vspace{-4mm}
\end{figure}

Now we explain how to chain the gadgets utilizing the subroutine epilogue in each of them. An adversary needs to craft a ``gadget stack'' and sends it along with the payload. Each gadget in the chain, except the last one,  has a corresponding gadget frame placed on the gadget stack. A gadget frame consists of several words of data that will be popped to the operand registers of the last \textit{POP} instruction in that gadget. The data provided by the gadget frame includes the address of the next gadget, which helps to jump to the next gadget after being popped. An example of a chain of three gadgets in ARMv8-M is presented in Figure \ref{graph::rop}. The payload contains the entry of Gadget 1 and two gadget frames corresponding to Gadget 1 and 2. To ensure the entry of Gadget 2 will be popped to \textit{LR}, the gadget frame for Gadget 1 contains three words since the second to last instruction in Gadget 1 pops three words from the stack. Two words before the entry of Gadget 2
are provided such that they will be popped to \textit{R4} and \textit{R5} instead. Similarly, the gadget frame for Gadget 2 provides two words of data which will be popped to \textit{R4} and \textit{PC} so that execution of Gadget 3 can be routed to start. 

\subsubsection{\textbf{Heap-based BOF Attacks}}
A heap-based BOF refers to a form of BOF exploitation in the heap area. 
As a SAM L11 project is linked with the GNU libc, the heap in SAM L11 is managed by the glibc allocator \cite{heap}. The glibc allocator manages free chunks in a doubly-linked list where each chunk contains the metadata of a forward pointer and a backward pointer pointing to the free chunk before and after it. 
A simple exploitation of heap-based BOF is to overwrite the function pointers stored on the heap to hijack the program control flow.
An adversary may also overwrite the metadata of a free chunk via overflowing an adjacent activated data chunk. By manipulating the pointers in the metadata, an adversary is able to corrupt arbitrary memory with arbitrary values \cite{xu2003transparent}.

\subsubsection{\textbf{Format String Attacks}}
\label{sec:fomat}
The format string exploits occur when a format function (e.g., \textit{printf()}) receives a format string input (the first argument of a format function), which contains more format specifiers (like \textit{\%s}, \textit{\%d}) than the arguments supplied. An adversary can send a well-crafted payload to a vulnerable format string function with number of specific format specifiers more than required arguments. This may eventually cause memory leakage or alteration in the stack, or even in an arbitrary readable/writable memory location. 

In SAM L11, an adversary is able to exploit format string vulnerabilities for reading values from the stack by sending a malicious text input containing more format specifiers. For example, by sending string \textit{``\%x \%x \%x"} to the vulnerable function illustrated in Listing \ref{list::fmtstr}, in which no arguments is provided to specifiers in the format string, three bytes of data will be printed in hexadecimal from the stack. 
However, reading or writing at an arbitrary memory location is unachievable in SAM L11 due to its particular memory addressing as shown in Figure \ref{graph::mem}. Such attacks require the target address be present in the payload. Adversaries who aim at the memory of SAM L11 will find that any address of the memory would contain at least one null byte. The format string function will terminate the process of parsing the payload when it hits the null byte.  

\lstset{style=mystyle}
\begin{lstlisting}[language=c, caption=Example of a vulnerable format string function, label = list::fmtstr] 
void fmt_str(char *input){
    printf(input);
    ... 
\end{lstlisting}

\subsubsection{\textbf{Attacks against Non-secure Callable (NSC) Functions}}
\label{sec::nsc}

The Non-secure software in the NSW may desire to use the Secure services in the SW. For the sake of such requirement, TrustZone-M provides the NSC memory region within the SW. Developers are able to define NSC functions in the NSC as the gateway to the SW. NSC functions are characterized with two features: (i) They can be called from the NSW; (ii) They have the privilege of accessing Secure resources since the NSC is a region within the SW. With such abilities, Non-secure software can call specific Secure services by first calling the corresponding NSC functions. The NSC functions then help to call the target Secure functions and pass the required arguments assigned by the Non-secure callers.

As the gateway to the SW, the implementation of the NSC software should be particularly cautious. According to the guidance from ARM \cite{armguidelines}, it is the common responsibility of hardware, toolchain, and software developers to implement the NSC software securely. Though some requirements are offered in the guidelines, since the hardware and toolchain vary from vendors to vendors, there is no off-the-shelf solution to implementing trusted NSC software. 

We identify two pitfalls that software developers may meet while programming the NSC functions.
The first pitfall is caused by the data arguments sent from the NSW. The toolchain of SAM L11 only helps to generate the Secure gateway veneer for NSC functions but leaves the function programming to the developers. Security related coding mistakes may present in the NSC functions as well and can be exploited by crafting Non-secure data inputs injected from NSW. Software exploits in the NSC region lead to a compromised SW. 
The second pitfall comes from the untrusted address input. When Non-secure software passes an pointer argument to the SW through NSC functions, the SW should ensure that the pointer is within the NSW. Otherwise, Secure functions may assist the NSW to read or write the Secure memory. The vulnerable NSC function illustrated in Listing \ref{list::nsc} can leak and corrupt the Secure memory content at arbitrary Secure addresses if the first and the third arguments are Secure addresses and the second argument is set to 1. The violation of the principle that ``Secure resources are not allowed to be accessed by the NSW'' severely harms the fundamental security of the TrustZone-M implementation.

\lstset{style=mystyle}
\begin{lstlisting}[language=c, caption=Example of a vulnerable NSC function, label = list::nsc] 
int NSC_func(int *a, int b, int *c){
    int *addr = a; int num = b; int *sum = c;
    for (int i = 0; i < num; i++){
        &sum += addr[i];}
    return &sum;}
\end{lstlisting}
\section{Evaluation}
\label{sec:evaluation}

\begin{table*}[htbp]
\centering
\caption{Sizes of payloads and experiment results in different attack scenarios.} \label{tab1}
\begin{tabular}{|c|c|l|}
\hline
\textbf{Attack Scenarios}& \textbf{Sizes of Payloads (byte)}& \textbf{\makecell{Experiment Results}}\\
\hline
Code injection& 256& 90.62s is spent on scanning the entry of the malicious code, which is then successfully executed.\\
\hline
ROP& 96& Crafted string is printed; 65 potential gadgets are found in a 4.14KB image.\\
\hline
Heap-based BOF& 24& The malicious code is successfully executed.\\
\hline
Format string exploits& 24& Five sequential bytes are read from the Non-secure and Secure stacks.\\
\hline
NSC-specific attacks& 24 \& 4& \tabincell{l}{3 of 5 demo projects contain vulnerable NSC functions; Five sequential bytes are read from \\the Secure stack; Secure memory content is printed.}\\
\hline
\end{tabular}
\vspace{-2mm}
\end{table*}

In this section, we evaluate the five software attacks described in Section \ref{sec:pitfall}. We are able to successfully perform these attacks against a TrustZone-M enabled MCU, SAM L11.

\subsection{Experiment Setup}
We use a laptop as the attacker to continually send text inputs to a SAM L11 Xplained Pro Evaluation Kit as the victim device. The laptop is connected with SAM L11 through a USB-to-UART adapter while an attacker may also inject malicious strings into an Internet connection of a SAM L11 based IoT device. 
In SAM L11, two UARTs are configured accordingly as a Non-secure peripheral and a Secure peripheral to receive inputs sent from the laptop to the NSW and SW respectively. For the first four attacks, we construct specific vulnerable functions in both Non-secure and Secure applications of SAM L11 and malicious payloads will be sent through the UARTs to trigger the attacks. The sizes of the payloads and experiment results are given in Table \ref{tab1}. 

\subsection{Experiment Results}
In the BOF-based code injection attack, 
we configure the stack to be executable, which is commonly configurable in TrustZone-M enabled MCUs.
We assemble the payload with components including a constant string, a malicious code, the entry of the malicious code (obtained via random brute-force scanning), and a NOP sled of 50 NOP instructions. The malicious code is designed to call a Non-secure print function and provide the address of the constant string as the argument to the print function. Our attack successes and the constant string is printed in the adversary's terminal.

As the proof-of-concept implementation of ROP, we craft three exploitable gadgets by splitting the assembly code of a program which prints content at a given memory location, into three code segments and appending a subroutine epilogue (i.e., \textit{POP PC}) at the end of each segment. These gadgets are pre-stored at different locations of the flash in advance. We craft a gadget stack to chain these gadgets and send it along with the payload to SAM L11.
As a result, the intended constant string is successfully printed in the adversary's terminal. A way to evaluate the feasibility of ROP against a certain program is to count up the occurrence of potential gadgets in the program. In fact, this process is equivalent to counting up the number of ``\textit{POP PC}'' and ``\textit{BX LR}'' instructions according to the definition of potential gadgets in Section \ref{sec:pitfall}. We take a basic NSW application image, which only initializes necessary peripherals, as an example and search all the subroutine epilogues in it. The Capstone disassembly engine \cite{capstone} is used to dissemble and search in the binary code. The size of the example image is 4.14KB with 1908 instructions in total. As a result, 49 ``\textit{POP PC}'' and 16 ``\textit{BX LR}'' are found in the image binary, representing 3.41\% of the whole image.

To launch the heap-based BOF attack, we first construct two adjacent data blocks on the heap of SAM L11 and a vulnerable \textit{strcpy()} function, which copies the input payload to a buffer in the first data block without checking its boundary. Our payload successfully triggers the BOF attack and overwrites a function pointer in next data block with the entry of a pre-injected malicious code. The malicious code is later executed when that function is called.

As we stated in Section \ref{sec:fomat}, an adversary can exploit the vulnerable format string function in SAM L11 to read out the stack content. The payload used is ``\%08x \%08x \%08x \%08x \%08x'' and we eventually read five sequential bytes from the stack via UARTs.

To verify the feasibility of NSC-specific attacks, we look into the example software projects provided by the vendor of SAM L11, five of which contain NSC software implementations.
We statically analyze the source code of these five NSC implementations and find three to be vulnerable. These three implementations share two vulnerable NSC functions as in Listing \ref{list::nscdemo}, where two of them contain the first function and the other contains the second function. The first vulnerable function is subject to the format string attack when it is called by the Non-secure software and the argument is a crafted format string input that can be controlled by an adversary. In our experiment,
we send ``\%08x \%08x \%08x \%08x \%08x'' as the payload and five sequential bytes from the Secure stack are eventually printed on the adversary's terminal.
The second function has an information leakage problem. We call this function in the NSW and the argument passed to this function is a Secure address. The Secure memory content at the target location is then printed.

\lstset{style=mystyle}
\begin{lstlisting}[language=c, caption=Vulnerable NSC functions in SAM L11 demo code, label = list::nscdemo] 
void __attribute__((cmse_nonsecure_entry)) nsc_secure_console_puts (char *string){
	non_secure_puts(string);}

void __attribute__((cmse_nonsecure_entry)) nsc_puts(uint8_t * string){
	printf("%s", string);}
\end{lstlisting}

\section{Discussion: Defense to Runtime Software Attacks against TrustZone-M}
\label{sec:defense}
The control flow integrity (CFI) may prevent runtime control-oriented attacks.
By monitoring the control flow of a program at runtime, it can detect unexpected control flow changes. 
In \cite{nyman2017cfi}, CFI is implemented for TrustZone-M to protect the NSW. The control flow graph (CFG) of a program is constructed by static or dynamic analysis of its code and is saved in a non-writable region of the NSW. Code instrumentation is performed so that the program jumps to a {\em branch monitor} before any control flow change in the original code.
The branch monitor refers to the CFG and monitors control flow changes at runtime. Before a function call, the correct return address is pushed on a {\em shadow stack} in the SW. Recall a function may be called by multiple functions, and return to different places at runtime. Since the CFG cannot tell the exact return address at runtime, the shadow stack in the SW is used to record the correct return address.

The CFI for protecting the control flow of the NSW is not sufficient for the overall system security. It can be observed from Table \ref{table::attack} that all software attacks may occur in all the three regions, i.e. the NSW, NSC and SWX, of TrustZone-M. 
Another issue of CFI is that it may not defeat the heap-based BOF or format string attacks if the control flow is not changed, but sensitive data is stolen or changed. 

For the overall security of TrustZone-M based IoT systems, the following strategies may be adopted. First, the SW shall not have the Internet connection in order to avoid remote exploits of SW software vulnerabilities. Second, the arguments sent from the NSW to an NSC function may be an address or application data. If it is an address, the NSC function must verify that it is not a Secure address before passing the address argument to the SW, since an NSW program shall not access the SW resources directly. 
If it is application data, input sanitization shall be carefully performed.
Third, security mechanisms including CFI and onboard executable space protection of TrustZone-M shall be applied to the NSW for control flow integrity.
Finally, secure coding, code review and penetration testing are critical to the overall system security and the best practice shall be adopted \cite{BISO:SecureCoding:2020}. Runtime software attacks may also be mitigated by programming the MCUs with security oriented languages such as Java \cite{ito2001making, stm32} and Rust \cite{Rust:Embedded:2020}.  
\section{Conclusion}
\label{sec:conclusion}

This paper gives the first systematic runtime software security analysis for TrustZone-M enabled IoT devices. We present possible pitfalls of TrustZone-M programming and present five potential software attacks against TrustZone-M, including the stack-based BOF attack for code injection, return-oriented programming, heap-based BOF attacks, format string attacks and attacks against NSC functions. We validate these attacks on a TrustZone-M enabled MCU, SAM L11. To defeat these attacks, we present guidelines for an overall system security of TrustZone-M enabled IoT devices.

\bibliographystyle{IEEEtran}
\bibliography{IEEEabrv,IEEEexample}

\begin{thebibliography}{10}
\providecommand{\url}[1]{#1}
\csname url@samestyle\endcsname
\providecommand{\newblock}{\relax}
\providecommand{\bibinfo}[2]{#2}
\providecommand{\BIBentrySTDinterwordspacing}{\spaceskip=0pt\relax}
\providecommand{\BIBentryALTinterwordstretchfactor}{4}
\providecommand{\BIBentryALTinterwordspacing}{\spaceskip=\fontdimen2\font plus
\BIBentryALTinterwordstretchfactor\fontdimen3\font minus
  \fontdimen4\font\relax}
\providecommand{\BIBforeignlanguage}[2]{{%
\expandafter\ifx\csname l@#1\endcsname\relax
\typeout{** WARNING: IEEEtran.bst: No hyphenation pattern has been}%
\typeout{** loaded for the language `#1'. Using the pattern for}%
\typeout{** the default language instead.}%
\else
\language=\csname l@#1\endcsname
\fi
#2}}
\providecommand{\BIBdecl}{\relax}
\BIBdecl

\bibitem{sonicwall}
\BIBentryALTinterwordspacing
(2020, Feb.) 2020 sonicwall cyber threat report: Threat actors pivot toward
  more targeted attacks, evasive exploits. SonicWall. [Online]. Available:
  \url{https://www.sonicwall.com/news/2020-sonicwall-cyber-threat-report/}
\BIBentrySTDinterwordspacing

\bibitem{owasp}
\BIBentryALTinterwordspacing
T.~O. I.~S. Team. (2018) Owasp internet of things project. [Online]. Available:
  \url{https://owasp.org/www-project-internet-of-things/}
\BIBentrySTDinterwordspacing

\bibitem{IoTnumber}
\BIBentryALTinterwordspacing
K.~Gyarmathy. (2020, Mar.) Comprehensive guide to iot statistics you need to
  know in 2020. [Online]. Available:
  \url{https://www.vxchnge.com/blog/iot-statistics}
\BIBentrySTDinterwordspacing

\bibitem{mohanty2018control}
A.~Mohanty, I.~Obaidat, F.~Yilmaz, and M.~Sridhar, ``Control-hijacking
  vulnerabilities in iot firmware: A brief survey,'' 2018.

\bibitem{haddadpajouh2019survey}
H.~HaddadPajouh, A.~Dehghantanha, R.~M. Parizi, M.~Aledhari, and H.~Karimipour,
  ``A survey on internet of things security: Requirements, challenges, and
  solutions,'' \emph{Internet of Things}, 2019.

\bibitem{english2019exploiting}
K.~V. English, I.~Obaidat, and M.~Sridhar, ``Exploiting memory corruption
  vulnerabilities in connman for iot devices,'' pp. 247--255, 2019.

\bibitem{nyman2017cfi}
T.~Nyman, J.-E. Ekberg, L.~Davi, and N.~Asokan, ``Cfi care: Hardware-supported
  call and return enforcement for commercial microcontrollers,'' in
  \emph{International Symposium on Research in Attacks, Intrusions, and
  Defenses}.\hskip 1em plus 0.5em minus 0.4em\relax Springer, 2017, pp.
  259--284.

\bibitem{tzm}
\BIBentryALTinterwordspacing
Trustzone for cortex-m. Arm. [Online]. Available:
  \url{https://www.arm.com/why-arm/technologies/trustzone-for-cortex-m}
\BIBentrySTDinterwordspacing

\bibitem{liu2019security}
L.~Liu, J.~Ma, C.~Zhang, T.~Chong, H.~Zhang, and Y.~Dong, ``Security software
  system design and implementation for microcontrollers based on trustzone,''
  \emph{DEStech Transactions on Computer Science and Engineering}, no. cisnrc,
  2019.

\bibitem{jung2020secure}
J.~Jung, J.~Cho, and B.~Lee, ``A secure platform for iot devices based on arm
  platform security architecture,'' in \emph{2020 14th International Conference
  on Ubiquitous Information Management and Communication (IMCOM)}.\hskip 1em
  plus 0.5em minus 0.4em\relax IEEE, 2020, pp. 1--4.

\bibitem{cerdeira2020sok}
D.~Cerdeira, N.~Santos, P.~Fonseca, and S.~Pinto, ``Sok: Understanding the
  prevailing security vulnerabilities in trustzone-assisted tee systems,'' in
  \emph{Proceedings of the IEEE Symposium on Security and Privacy (S\&P), San
  Francisco, CA, USA}, 2020, pp. 18--20.

\bibitem{iannillo2019proposal}
A.~K. Iannillo and R.~State, ``A proposal for security assessment of
  trustzone-m based software,'' in \emph{2019 IEEE International Symposium on
  Software Reliability Engineering Workshops (ISSREW)}.\hskip 1em plus 0.5em
  minus 0.4em\relax IEEE, 2019, pp. 126--127.

\bibitem{song2019sok}
D.~Song, J.~Lettner, P.~Rajasekaran, Y.~Na, S.~Volckaert, P.~Larsen, and
  M.~Franz, ``Sok: sanitizing for security,'' pp. 1275--1295, 2019.

\bibitem{rop}
\BIBentryALTinterwordspacing
Return oriented programming (arm32). Azeria Labs. [Online]. Available:
  \url{https://azeria-labs.com/return-oriented-programming-arm32/}
\BIBentrySTDinterwordspacing

\bibitem{nxbit}
\BIBentryALTinterwordspacing
Nx bits - microsoft wiki - fandom. Microsoft. [Online]. Available:
  \url{https://microsoft.fandom.com/wiki/NX bit}
\BIBentrySTDinterwordspacing

\bibitem{heap}
\BIBentryALTinterwordspacing
Arm heap exploitation. Azeria Labs. [Online]. Available:
  \url{https://azeria-labs.com/heap-exploitation-part-1-understanding-the-glibc-heap-implementation/}
\BIBentrySTDinterwordspacing

\bibitem{xu2003transparent}
J.~Xu, Z.~Kalbarczyk, and R.~K. Iyer, ``Transparent runtime randomization for
  security,'' pp. 260--269, 2003.

\bibitem{armguidelines}
\BIBentryALTinterwordspacing
Armv8-m secure software guidelines. Arm. [Online]. Available:
  \url{https://developer.arm.com/docs/100720/0200/secure-software-guidelines}
\BIBentrySTDinterwordspacing

\bibitem{capstone}
\BIBentryALTinterwordspacing
The ultimate disassembly framework – capstone – the ultimate disassembler.
  [Online]. Available: \url{http://www.capstone-engine.org/}
\BIBentrySTDinterwordspacing

\bibitem{BISO:SecureCoding:2020}
\BIBentryALTinterwordspacing
{Berkeley Information Security Office}. (2020) Secure coding practice
  guidelines. [Online]. Available:
  \url{https://security.berkeley.edu/secure-coding-practice-guidelines}
\BIBentrySTDinterwordspacing

\bibitem{ito2001making}
S.~A. Ito, L.~Carro, and R.~P. Jacobi, ``Making java work for microcontroller
  applications,'' \emph{IEEE Design \& Test of Computers}, vol.~18, no.~5, pp.
  100--110, 2001.

\bibitem{stm32}
\BIBentryALTinterwordspacing
Stm32 ide. STMicroelectronics. [Online]. Available:
  \url{https://www.st.com/en/development-tools/stm32-ides.html}
\BIBentrySTDinterwordspacing

\bibitem{Rust:Embedded:2020}
\BIBentryALTinterwordspacing
(2020) Rust: Embedded devices. [Online]. Available:
  \url{https://www.rust-lang.org/what/embedded}
\BIBentrySTDinterwordspacing

\end{thebibliography}
 
\end{document}